\documentclass[
amssymb,aps,
prd,nofootinbib,floatfix,12pt,superscriptaddress,
]{revtex4}%
\usepackage[T1]{fontenc}
\usepackage[utf8]{inputenc}
\usepackage{textcomp}
\usepackage{amssymb}
\usepackage{amsmath}
\usepackage{epsfig}
\usepackage{graphicx}
\usepackage{esint}
\usepackage{comment}
\usepackage{color}
\PassOptionsToPackage{normalem}{ulem}
\usepackage{ulem}
\usepackage{slashed}
\usepackage{caption}
\usepackage{amsmath,amssymb,epsf}
\usepackage{graphicx,wrapfig,color}
\usepackage{subfig}
\usepackage{setspace}
\usepackage{makecell}
\usepackage{tabularx}
\usepackage{booktabs}
\usepackage[tablename=Table]{caption}
\usepackage{float}
\usepackage{placeins}
\usepackage[unicode=true,pdfusetitle,
 bookmarks=true,bookmarksnumbered=false,bookmarksopen=false,
 breaklinks=false,pdfborder={0 0 0},backref=false,colorlinks=true,citecolor=blue,urlcolor=violet 
]{hyperref}
\DeclareMathOperator{\sech}{sech}
\usepackage{xcolor}
\usepackage{footnote}

\newcommand{\orcid}[1]{\href{https://orcid.org/#1}{\includegraphics[width=10pt]{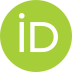}}}
\newcommand{\be}{\begin{equation}}
\newcommand{\ee}{\end{equation}}
\newcommand{\bea}{\begin{eqnarray}}
\newcommand{\eea}{\end{eqnarray}}

\newcommand{\ba}{\begin{array}}
\newcommand{\ea}{\end{array}}
\newcommand{\bd}{\begin{displaymath}}
\newcommand{\ed}{\end{displaymath}}

\def\gev{{\rm \,Ge\kern-0.125em V}}
\def\tev{{\rm \,Te\kern-0.125em V}}

\def\th13 {\theta_{13}}

\begin{document}
\title{Study of Goldstone Inflation in the domain of Einstein-Gauss-Bonnet gravity}
\author{Hussain Ahmed Khan\orcid{0000-0003-1166-5190}}
\email{hussainahmed@ctp-jamia.res.in}
\affiliation{Centre For Theoretical Physics, Jamia Millia Islamia, New Delhi-110025, India.}

\author{Yogesh\orcid{0000-0002-7638-3082}}
\email{yogesh@ctp-jamia.res.in}
\affiliation{Centre For Theoretical Physics, Jamia Millia Islamia, New Delhi-110025, India.}

\begin{abstract}
\noindent
 Realizing the inflationary epoch driven by a  pseudo-Nambu Goldstone boson (pNGB) could ensure the coveted flatness, and the sub-Planckian scales related to the dynamics of the paradigm. In this work, we have taken the most general form of such a scenario: Goldstone Inflation, proposed in \cite{croon}, and studied the model in Einstein-Gauss-Bonnet gravity. Natural inflation, which is a limiting case of this model, is also studied here. The specific form of the EGB coupling gives ample opportunity to study the rich  phenomenology associated with inflation as well as the reheating epoch. Predicted values of the inflationary observables, tensor to scalar ratio ($r$), and spectral index ($n_s$) are in good agreement with the recent observations from $Planck'18$ \cite{Planck2018}. Thus, in the framework of EGB one can resurrect the model, which otherwise needs quite a bit of fine tuning or diversion from the canonical domain as studied in \cite{Bhattacharya:2018xlw}, to survive in the standard cold inflationary scenario. Finally, the era of reheating is studied for different choices of model parameters.

\end{abstract}
\maketitle

\newpage

\section{Introduction}
From the onset of the introduction of inflationary paradigm in the seminal paper by Alan Guth\cite{guth}, to resolve some of the hot big bang cosmology puzzles(reviews can be found in \cite{Liddle},\cite{cmbinflate,Linde:2005ht}), it has become one of the primary fields of research at the interface of particle physics and cosmology(readers are advised to go through \cite{Linde:1983gd}-\cite{mukha81} for important early works). In the last few decades, cosmology has advanced tremendously in the observational sector, becoming a part of precision physics. The stupendous advancement in the observations made it possible to constrain different theoretical predictions from real data. Most of the so called text book models of inflation are being ruled out by CMB observations from WMAP \cite{WMAP9} and Planck \cite{PlanckXX}, at least in the standard cold inflationary scenario. After the final observational data reported by Planck\cite{Planck2018}, it can at least be said, data prefers sub Planckian small field models of inflation, which, in a sense, is satisfactory from the idea of the effective field theory.

The Goldstone inflationary model was first proposed in \cite{croon}, from the idea of minimal composite Higgs model  \cite{csaki,contino}. Precedingly the Natural inflation model was proposed \cite{freese} (also ref. to \cite{Adams:1992bn,Kim:2004rp,delaFuente:2014aca,Jensen:1986nf,Freese:2014nla}). This model used an axion as the inflaton, which is the Goldstone of a spontaneously broken Peace-Quinn symmetry. With a breaking scale of $10 M_{pl}$ or higher, the model still lies in the $2-\sigma$ allowed region. But this has its issues because the dynamics of the effective field theory could get compromised by the effects of quantum gravity, which play a significant role in the super-planckian domain. Generally quantum gravity does not conserve global symmetry, so to have a breaking scale that is super-planckian in the case of vanilla natural inflation, is philosophically not viable. Even though it has been studied that with sufficient fine tuning, Goldstone inflation can be rescued with a sub-planckian breaking scale, but in the light of new data from the {\it Planck’18 } it does survive even in the canonical domain. Due to the fact that Goldstone model of inflation is derived from minimal composite Higgs model, noncanocial origin of the inflationary dynamics is expected \cite{Bhattacharya:2018xlw}. To explain the super-planckian breaking scale different models have been proposed, namely, the extranatural inflation \cite{nima}, hybrid axion models \cite{lindehyb, kim}, N-flation \cite{nflation1} -\cite{nflation3}, axion monodromy \cite{axionmono}, and other pseudo natural inflationary models in supersymmetry \cite{pngb}. A large amount of fine tuning and extra dimensions are required to save these models of inflation.

The resurrection of the Goldstone model of inflation and Natural inflation, which is a limiting case of Goldstone inflation, has been tried in the standard inflationary scenario. But so far this pursuit has been unsuccessful for sub-Planckian breaking scales ($f$), as the values of the inflationary observables obtained are not favoured experimentally from {\it Planck’18}.  To save these inflationary models in the standard case the breaking scale has to be super-Planckian. 

It has been studied that in modified theories of gravity, the restrictions on inflationary models can be made less stringent \cite{Kallosh:2013pby,predictions,Barvinsky:1994hx,Cervantes-Cota1995,BezrukovShaposhnikov,EOPV2014,Koshelev:2020xby}., which was otherwise not possible in the standard theory of General Relativity. We will therefore consider the validity of Goldstone inflation and Natural inflation in the domain Einstein-Gauss-Bonnet gravity, in the sub-planckian regime. Here a term, namely, the Gauss-Bonnet term, is added to the Einstein-Hilbert action, which does not make any difference to the equations of motion as it is a total derivative. But if we couple it with a function of the scalar field $\phi$ as $\xi (\phi)$, it does become dynamically of substance. Another important thing to note here is that that Gauss-Bonnet term is an inherent quantum correction, in the domain of string theory, to the Einstein-Hilbert action.
A number of inflationary models have been studied in the EGB scenario \cite{Soda2008,Guo:2009uk,Guo,Jiang:2013gza,Koh,vandeBruck:2015gjd,Pozdeeva:2020apf,Pozdeeva:2016cja,JoseMathew,vandeBruck:2016xvt,Koh:2016abf,Nozari:2017rta,Armaleo:2017lgr,Chakraborty:2018scm,Yi:2018gse,Odintsov:2018,Fomin:2019yls,Kleidis:2019ywv,Rashidi:2020wwg,Odintsov:2020sqy,Pozdeeva:2020shl,Kawai:2021bye,Kawai:2017kqt}. It has been seen that in the EGB domain, while the value of spectral index $n_{s}$ remains unchanged, the tensor-to-scalar ratio becomes small, consistent with observational constraints.The most actively studied models with GB coupling involves the function $\xi$ inversely proportional to
the scalar field potential~\cite{Guo:2009uk,Guo,Jiang:2013gza,Koh:2016abf,Yi:2018gse,Odintsov:2018,Kleidis:2019ywv,Rashidi:2020wwg,Pozdeeva:2020shl}. In this work we choose a different form of the coupling, which has a $\tanh$ term.  The specific choice of coupling has found suitable for the study of Goldstone and Natural inflation.

The rest of the paper is organised as follows. In section \ref{sec2} we will discuss some of the basic yet essential aspects of inflation in the EGB framework. In section \ref{effectivepotential}, we will discuss the formalism, and the slow roll approximation of the effective potential in the framework of EGB. In section \ref{Goldstone Inflation in EGB}, the inflationary observables are calculated for Goldstone Inflation. In section \ref{Natural Inflation}, we have carried out the analysis for Natural Inflation. In section \ref{reheatingegb}, we analyzed the reheating epoch for the mentioned potential in section  \ref{Goldstone Inflation in EGB} and \ref{Natural Inflation}. Finally in section \ref{egbconclusion}, we conclude our work. 

\section{Inflation in EGB}
\label{sec2}
We will be considering a modified model of gravity with the Gauss-Bonnet term (we use the  reduced Planck mass, $M_p=1$),~\cite{Pozdeeva:2020apf,Guo:2009uk,Guo}:
\begin{equation}
\label{action1}
S=\int d^4x\sqrt{-g}\left[UR-\frac{g^{\mu\nu}}{2}\partial_\mu\phi\partial_\nu\phi-V(\phi)-\frac{\xi(\phi)}{2}\mathcal{G}\right],
\end{equation}
here, $V(\phi)$, and $\xi(\phi)$ are differentiable functions, $U$ is a positive constant, and $\mathcal{G}$ is the Gauss-Bonnet term given as,  $$\mathcal{G}=R_{\mu\nu\rho\sigma}R^{\mu\nu\rho\sigma}-4R_{\mu\nu}R^{\mu\nu}+R^2$$

The following set of equations, in a spatially flat Friedmann universe, can be derived by varying the action (\ref{action1}) with respect to the scalar field $\phi$, ~\cite{vandeBruck:2015gjd,Pozdeeva:2019agu}:
\begin{eqnarray}
               12UH^2 &=& \dot\phi^2+2V+24\dot{\xi}H^3,  \label{Equ0} \\
                4U\dot{H} &=& {}-\dot\phi^2+4\ddot{\xi}H^2+4\dot{\xi}H\left(2\dot{H}-H^2\right), \label{EquH}\\
                \ddot{\phi}&=&{}-3H\dot{\phi}-V'-12\xi'H^2\left(\dot{H}+H^2\right), \label{Equphi}
\end{eqnarray}
here, the dots and prime represent the derivatives taken with respect to the cosmic time $t$ and the scalar field $\phi$, and $H=\dot{a}/a$ is the Hubble parameter, where $a$ is the scale factor.

We consider the slow roll parameters as given in Refs.~\cite{Guo,vandeBruck:2015gjd}:
\begin{eqnarray}
  \epsilon_1 &=&{}-\frac{\dot{H}}{H^2}={}-\frac{d\ln(H)}{dN},\qquad \epsilon_{i+1}= \frac{d\ln|\epsilon_i|}{dN},\quad i\geqslant 1, \\
  \delta_1&=& \frac{2}{U}H\dot{\xi}=\frac{2}{U}H^2\xi'\frac{d{\phi}}{dN},\qquad \delta_{i+1}=\frac{d\ln|\delta_i|}{dN},\quad i\geqslant 1,
\end{eqnarray}
here we are considering ${d}/{dt}=H\, {d}/{dN}$. 

The slow roll approximation will then require, $$ |\epsilon_i| \ll 1 \; \text{and} \;  |\delta_i| \ll 1  
$$

Eqs.~(\ref{Equ0})--(\ref{Equphi}) can be simplified using the slow roll conditions $\epsilon_1 \ll 1, \; \epsilon_2 \ll 1 \; \delta_1 \ll 1 \; \text{and} \; \delta_2 \ll 1. $
Therefore we obtain,
\begin{equation}
\label{delta2equ}
\delta_2=\frac{\dot\delta_1}{H\delta_1}=\frac{2\ddot\xi}{U\delta_1}-\epsilon_1,
\end{equation}
and we have $|\ddot\xi|\ll |H\dot\xi|$ from $|\delta_2|\ll 1$ and $|\epsilon_1|\ll 1$.

Using the conditions $|\delta_1|\ll 1$ and $|\delta_2|\ll 1$, in Eqs.~(\ref{Equ0}) and (\ref{EquH}), we get:
\begin{eqnarray}
               &&12UH^2 \simeq \dot\phi^2+2V,  \label{Equ0slr1} \\
                &&4U\dot{H} \simeq{}-\dot\phi^2-4\dot{\xi}H^3={}-\dot{\phi}\left(\dot{\phi}+4\xi'H^3\right). \label{EquHslr1}
\end{eqnarray}
Using,
\begin{equation*}
\epsilon_1={}-\frac{\dot H}{H^2}\simeq \frac{\dot\phi^2}{3(\dot\phi^2+2V)}+\frac{1}{2}\delta_1\ll 1,
\end{equation*}
we obtain $\dot\phi^2 \ll 2V$, and Eq.~(\ref{Equ0slr1}) takes the following form
\begin{equation}
\label{Equ0slr2}
6UH^2 \simeq V.
\end{equation}
Differentiating the above equation with respect to time and making use of Eq.~(\ref{EquHslr1}), we obtain
\begin{equation}
\label{Equphislr1}
    \dot{\phi}\simeq{}-\frac{V'}{3H}-4\xi'H^3.
\end{equation}
Substituting (\ref{Equphislr1}) into Eq.~(\ref{Equphi}), we get $|\ddot{\phi}|\simeq|12\xi'H^2\dot{H}|\ll|12\xi'H^4| $.

Therefore, the slow-roll conditions give the following results:
 \begin{equation*}
 \dot\phi^2\ll V, \quad |\ddot{\phi}|\ll |12\xi'H^4|,\quad 2|\dot{\xi}|H\ll U,\quad |\ddot{\xi}|\ll|\dot{\xi}|H\,,
 \end{equation*}
 and the leading order equations in the slow-roll approximation will be:
 \begin{eqnarray}
               H^2&\simeq&\frac{V}{6U}\,, \label{Equ0lo}\\
               \dot{H}&\simeq&{}-\frac{\dot\phi^2}{4U}-\frac{\dot{\xi}H^3}{U}\,, \label{EquHlo}\\
                \dot{\phi}&\simeq&{}-\frac{V'+12\xi'H^4}{3H}. \label{Equphilo}
\end{eqnarray}

We do the inflationary analysis using the effective potential formalism, which we will discuss in the following section. 
\section{The effective potential}
\label{effectivepotential}
\subsection*{The slow-roll approximation}
\label{slow-roll approximation}
To analyze the stability of de Sitter solutions in model (\ref{action1}) the effective potential has been proposed in Ref.~\cite{Pozdeeva:2019agu}:
\begin{equation}
\label{Veff}
V_{eff}(\phi)={}-\frac{U^2}{V(\phi)}+\frac{1}{3}\xi(\phi).
\end{equation}

For $V(\phi)\equiv 0$, the potential is undefined, but for such a case, inflationary scenarios are always unstable.~\cite{Hikmawan:2015rze} (see also~\cite{Chakraborty:2018scm}).
In this work,
we will be considering inflationary scenarios with positive potentials only: $V(\phi)> 0$ during inflation.
The effective potential characterizes existence and stability of de Sitter solutions completely.
It is however not enough to completely characterize quasi-de Sitter inflationary stage, and the potential
$V(\phi)$ enters into equations for the inflationary parameters as well. But we will
keep the effective potential in the corresponding formulae, 
as it would be helpful in our analysis. 

Using Eqs.~(\ref{EquHlo}) and (\ref{Equphilo}), we get that the functions $H(N)$ and $\phi(N)$ satisfy the following leading order equations:
\begin{eqnarray}
               \frac{d{H}}{dN}&\simeq&{}-\frac{H}{U}V'V_{eff}'\,, \label{EquHloN}\\
               \frac{d{\phi}}{dN}&\simeq&{}-2\frac{V}{U}V_{eff}'. \label{EquphiloN}
\end{eqnarray}

In terms of the effective potential the slow-roll parameters are as follows:
\begin{equation}
    \epsilon_1={}-\frac{1}{2}\frac{d\ln(V)}{dN}=\frac{V'}{U}V_{eff}'\,,
    \label{EGBsr1}
    \end{equation}
\begin{equation}
    \epsilon_2={}-\frac{2V}{U}V_{eff}'\left[\frac{V''}{V'}+\frac{V_{eff}''}{V_{eff}'}\right]
    ={}-\frac{2V}{U}V_{eff}'\left[\ln(V'V_{eff}')\right]'\,,
    \label{EGBsr2}
\end{equation}
\begin{equation}
  \delta_1= {}-\frac{2V^2}{3U^3}\xi'V_{eff}'\,,
  \label{EGBsr3}
  \end{equation}
\begin{equation}
\begin{split}
\delta_2=& {}-\frac{2V}{U}V_{eff}'\left[2\frac{V'}{V}+\frac{V_{eff}''}{V_{eff}'}+\frac{\xi''}{\xi'}\right]\\
=&{}-\frac{2V}{U}V_{eff}'\left[\ln(V^2\xi'V_{eff}')\right]'.
\end{split}
\label{slrVeffd}
\end{equation}

So, $|\epsilon_1|\ll 1$ and $|\delta_1|\ll 1$ if $V_{eff}'$ is small enough. It allows us to use the effective potential for construction of the inflationary scenarios in models with the GB term.

Using the known formulae~\cite{Guo,Koh:2016abf} for  the tensor-to-scalar ratio $r$ and the spectral index $n_s$, we obtain:
\begin{equation}
\label{rVeff}
 r=8|2\epsilon_1-\delta_1|
\end{equation}
\begin{equation}
\label{nsVeff}
\begin{split}
   n_s=&1-2\epsilon_1-\frac{2\epsilon_1\epsilon_2-\delta_1\delta_2}{2\epsilon_1-\delta_1}
\end{split}
\end{equation}

A standard way to reconstruct inflationary models~\cite{Mukhanov:2013tua,Koh:2016abf,Pozdeeva:2020shl} includes the assumption of explicit form of the inflationary parameter $n_s$ and $r$ as functions of $N$.  

The expression for amplitude $A_s$ in the leading order approximation is~\cite{vandeBruck:2015gjd}:
\begin{equation}
\label{As}
A_s\approx\frac{H^2}{\pi^2 U r}\approx\frac{V}{6\pi^2 U^2r}.
\end{equation}

In the slow-roll approximation, the e-folding number $N$ can be presented as the following function of $\phi$:
\begin{equation}
\label{N1}
N(\phi)=\int\limits^{\phi}_{\phi_{end}}\frac{U}{2VV'_{eff}}d\phi.
\end{equation}
 To get a suitable inflationary scenario we calculate inflationary parameters for $55\leq N\leq 75$, and compare them with the observation data~\cite{Planck2018}.
\section{Goldstone Inflation in EGB }
\label{Goldstone Inflation in EGB}
\noindent
In Goldstone inflation, the form of the potential is given as,
\begin{equation}
V(\phi) = V_0 \bigg[ C_{\Lambda}+ \alpha \cos \left(\frac{\phi}{f}\right) + \beta \sin^2 \left(\frac{\phi}{f}\right) \bigg]
\label{GIpotential}
\end{equation}
We choose the EGB coupling of the following form \cite{Kawai:2021edk},
\begin{equation}
\xi (\phi) = \frac{\xi_1}{V_0} \tanh(\xi_{2}\phi),
\end{equation}
where we have normalized the coupling by $V_0$, which makes the slow roll parameters independent of $V_0$. The scale of inflation can be fixed from the definition of $A_s$. Another advantage of this particular choice of EGB coupling is that it ensures that the shape of the effective potential is in concordance with the conditions under which inflation can take place.

In this case the slow-roll parameters can written using Eqs. (\ref{EGBsr1}),(\ref{EGBsr2}), (\ref{EGBsr3}) and (\ref{slrVeffd}) as:

\begin{footnotesize}
\begin{equation}
\epsilon_1 = \frac{1}{3 f U}  \Bigg\{ \left( - \alpha + 2 \beta \cos \left( \frac{\phi}{f} \right) \right)  \sin \left( \frac{\phi}{f} \right) \Bigg[ \xi_1 \xi_2 \sech^2 (\xi_2 \phi) -\frac{3 U^2 \left( \alpha-2 \beta \cos \left(\frac{\phi}{f} \right) \right) \sin \left( \frac{\phi}{f}\right)  }{f \left( C_{\Lambda}+ \alpha \cos \left( \frac{\phi}{f}\right) +\beta \sin^2 \left( \frac{\phi}{f}\right) \right)^2}    \Bigg]  \Bigg\}
\label{giepsilon1}
\end{equation}
\begin{align}
\epsilon_2 = &- \frac{1}{3 U^2} \left( C_{\Lambda}+\alpha \cos\left( \frac{\phi}{f} \right)  + \beta \sin^2 \left( \frac{\phi}{f} \right) \right)  \bigg( \xi_1 \xi_2 \sech^2(\xi_2 \phi) - \frac{A}{f~B^2} \bigg) \nonumber\\& \Bigg[ \frac{\left(\alpha \cos \left(  \frac{\phi}{f}\right)-2 \beta \cos \left( \frac{2 \phi}{f} \right) \right)  \csc \left( \frac{\phi}{f} \right)}{f \left(\alpha- 2 \beta \cos \left(\frac{\phi}{f} \right)  \right)} - \frac{\frac{C}{f^2~B^2}+ \frac{D}{f^2~B^3}+ 2 \xi_1 \xi_2^2 \sech^2 (\xi_2 \phi) \tanh (\xi_2 \phi)}{\xi_1 \xi_2 \sech^2(\xi_2 \phi) - \frac{A}{f~B^2}}    \Bigg]
    \end{align}
\label{epsilon22}
where,  $A= 3 U^2 \left( \alpha - 2 \beta \cos \left( \frac{\phi}{f} \right) \right) \sin \left( \frac{\phi}{f} \right), B =  C_{\Lambda} + \alpha \cos \left(\frac{\phi}{f} \right)+ \beta \sin^2 \left( \frac{\phi}{f} \right) , \\ C= 3 U^2 \left( \alpha \cos \left( \frac{\phi}{f} \right)- 2 \beta \cos \left( \frac{2 \phi}{f} \right) \right)$ and $D= 6 U^2 \left( \alpha \sin \left( \frac{\phi}{f} \right) - \beta \sin \left(\frac{2 \phi}{f} \right) \right)^2$
\begin{align}
\delta_1 =&- \frac{1}{9 f U^3} \Bigg\{ 2~ \xi_1~ \xi_2 \sech^2 \left( \frac{\phi}{f} \right) \Bigg[ \frac{1}{4} f ~\xi_1 ~\xi_2 \left( 2~C_{\Lambda} +\beta + 2 \alpha \cos \left(  \frac{\phi}{f}\right) - \beta \cos \left(  \frac{2 \phi}{f}\right)  \right)^2 \times  \nonumber\\&  \sech^2 \left( \xi_2 \phi \right) - 3 U^2 \left( \alpha- 2 \beta \cos \left( \frac{\phi}{f} \right) \sin \left( \frac{\phi}{f} \right) \right)    \Bigg]  \Bigg\}
\label{gidelta1}
\end{align}
\begin{align}
\delta_2 =& \frac{1}{3 f^2 U \left(C_{\Lambda} +\alpha \cos \left( \frac{\phi}{f}\right) + \beta \sin^2 \left( \frac{\phi}{f} \right)  \right)} \Bigg\{  2 f \xi_1 \xi_2  \left( 2 C_{\Lambda} + \beta + 2 \alpha \cos \left( \frac{\phi}{f}\right) - \beta \cos \left(  \frac{2 \phi}{f}  \right)   \right) \nonumber\\ & \sech^3 (\xi_2 \phi)  \bigg[ \cosh (\xi_2 \phi) \left( \alpha \sin \left( \frac{\phi}{f} \right) - \beta \sin \left( \frac{2 \phi}{f} \right) \right) +f \xi_2 \left( 2 C_{\Lambda} + \beta + 2 \alpha \cos \left( \frac{\phi}{f} \right) - \beta \cos \left( \frac{2 \phi}{f} \right)  \right) \nonumber\\ & \sinh (\xi_2 \phi) \bigg] + 6 U^2 \left[ \alpha \cos \left( \frac{\phi}{f} \right)- 2 \beta \cos \left( \frac{2 \phi}{f}  \right) - 2 f \xi_2  \left( \alpha- 2 \beta \cos \left( \frac{\phi}{f} \right) \right) \sin \left( \frac{\phi}{f} \right) \tanh (\xi_2 \phi) \right]  \Bigg\}
\label{gidelta2}
\end{align}
\end{footnotesize}

Using Eq. (\ref{N1}) we can write the expression of number of e-folds for the Goldstone inflation as:
\begin{footnotesize}
\begin{align}
N = \int\limits^{\phi}_{\phi_{end}}-\frac{3 f U \left( C_{\Lambda}+ \alpha \cos \left( \frac{\phi}{f} \right) + \beta \sin^2 \left( \frac{\phi}{f} \right)  \right)}{- \frac{1}{2} f \xi_1 \xi_2 \left( 2 C_{\Lambda} + \beta + 2 \alpha \cos \left( \frac{\phi}{f} \right) - \beta \cos \left( \frac{ 2\phi}{f} \right) \right)^2 \sech^2(\xi_2 \phi) + 6 U^2  \left( \alpha- 2 \beta \cos \left( \frac{\phi}{f} \right) \right) \sin \left( \frac{\phi}{f} \right)}
\label{giefolding}
\end{align}
\end{footnotesize}
To check the viability of Goldstone inflation in the framework of EGB gravity we calculate the inflationary observables, $n_{s}$ and $r$, using (\ref{nsVeff}) and (\ref{rVeff}). Following \cite{croon},  we consider the potential parameters $C_{\Lambda}= \alpha =1$ and $\beta = \frac{1}{2}$ .  For a particular choice of the parameters $U$, $\xi_{1}$ and $\xi_{2}$, we calculate the inflationary observations for different values of $f$. We fixed $\xi_1=13.3$ and $\xi_2=2.3$, which ensures the flatness of the potential. However, due to the non-trivial nature of the analysis, it is difficult to solve Eq. (\ref{giefolding}) analytically. So, here we use numerical techniques to evaluate the inflationary observables. Using Eq. (\ref{giepsilon1} - \ref{gidelta2}), we compute $r$ and $n_s$, for different combinations of $U$ and $f$, mentioned in the tables \ref{tab1},\ref{tab2}.

{\bf \underline {Case 1}} : U=0.5\\
For this case we take $U=0.5$, and calculate the inflationary observables $r$ and $n_s$ for three different values of $f$.

\begin{center}
\begin{table*}[!ht]
\begin{center}
\begin{tabular}{|c|c|c|c|c|c|c|}
\hline
   & \multicolumn{2}{c|}{$f=0.5$} & \multicolumn{2}{c|}{$f=0.7$} & \multicolumn{2}{c|}{$f=0.9$} \\ \hline
$N$  & $n_s$             & $r$                                           & $n_s$             & $r$                                         & $n_s$             & $r$               \\ \hline
55 & 0.9453          & $1.5226 \times 10^{-6} $           & 0.9454         & $5.8390 \times 10^{-6}$       & 0.9445        & $1.6577 \times 10^{-5}$          \\ \hline
65 & 0.9540         & $9.0211 \times 10^{-6}$             & 0.9532          & $3.6696 \times 10^{-6}$       & 0.9530        & $1.0282 \times 10^{-5}$        \\ \hline
75 & 0.9601         & $5.8974 \times 10^{-7}$            & 0.9594          & $2.3890 \times 10^{-6}$        & 0.9598       & $6.3976 \times 10^{-6}$         \\ \hline
\end{tabular}
\end{center}
\caption{Values of the inflationary parameters $r$ and $n_{s}$ for different values of $f$ and number of e-folds $N$. The observables are in good agreement with the  $Planck'18$ \cite{Planck2018} }.
\label{tab1}
\end{table*}
\end{center}

\begin{figure}[!htb]
\centering
\includegraphics[scale=0.45]{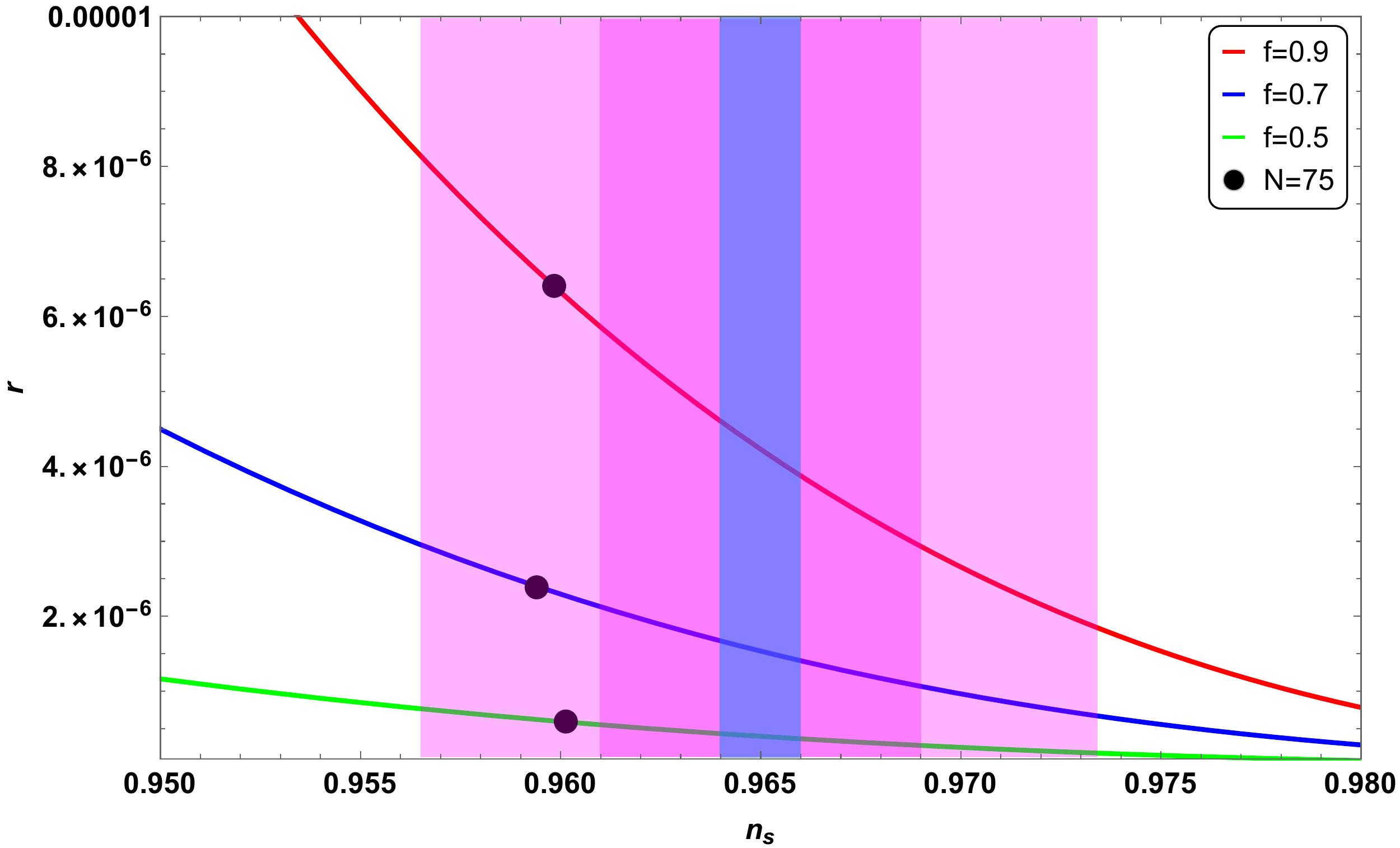}
\caption{Plots of $r$ and $n_s$, for $U=0.5$, for Goldstone inflation. The light pink shaded region corresponds to $2-\sigma$ and dark pink shaded region corresponds to $1-\sigma$ bounds on $n_s$ from {\it Planck'18}. The deep blue shaded region corresponds to the $1-\sigma$ bounds of future CMB observations \cite{Euclid,PRISM} keeping the same central value.   }
\label{rnsU05}
\end{figure}

{\bf \underline {Case 2}} : U=0.005\\
In this case we check the values of the inflationary observables for $U=0.005$, and keeping the other parameters constant as mentioned above. 
\begin{center}
\begin{table*}[!htb]
\begin{center}
\begin{tabular}{|c|c|c|c|c|c|c|}
\hline
   & \multicolumn{2}{c|}{$f=0.5$} & \multicolumn{2}{c|}{$f=0.7$} & \multicolumn{2}{c|}{$f=0.9$} \\ \hline
$N$  & $n_s$             & $r$            & $n_s$             & $r$           & $n_s$             & $r$               \\ \hline
55 & 0.9513         & 0.01658          &0.9572       & 0.04065      & 0.9605       & 0.06437         \\ \hline
65 & 0.9576         & 0.01069         & 0.9626       & 0.02850      & 0.9658        & 0.04767        \\ \hline
75 & 0.9624         & 0.00727         & 0.9666       & 0.02066      &0.9696       & 0.03642         \\ \hline
\end{tabular}
\end{center}
\caption{Values of the inflationary parameters $r$ and $n_{s}$ for different values of $f$ and number of e-folds $N$. The observables are in good agreement with the $Planck'18$ \cite{Planck2018} }.
\label{tab2}
\end{table*}
\end{center}

\begin{figure}[!htb]
\centering
\includegraphics[scale=0.44]{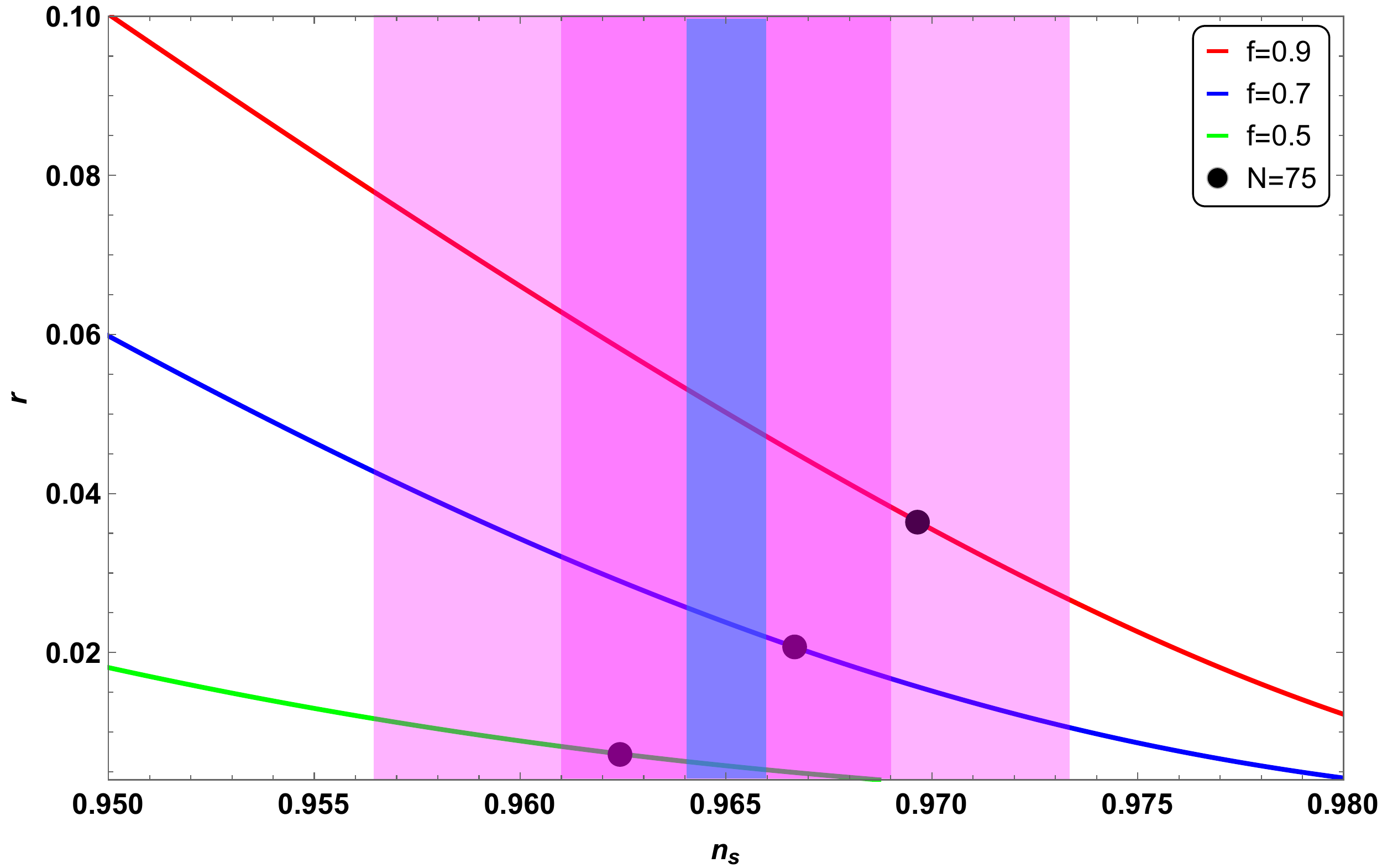}
\caption{Plots of $r$ and $n_s$, for $U=0.005$, for Goldstone inflation.  The light pink shaded region corresponds to $2-\sigma$ and dark pink shaded region corresponds to $1-\sigma$ bounds on $n_s$ from {\it Planck'18}. The deep blue shaded region corresponds to the $1-\sigma$ bounds of future CMB observations  \cite{Euclid,PRISM} keeping the same central value.   }
\label{rnsU005}
\end{figure}

\newpage

\section{Natural Inflation }
\label{Natural Inflation}
\noindent
Setting $C_{\Lambda}= \alpha=1$ and $\beta =0$ in Eq. (\ref{GIpotential}), we obtain the form of Natural inflation \cite{freese}. For sub-Planckian value of $f$, it has been extensively studied in the literature that in standard cold inflationary scenario this model is not congruent with recent observation. So, it will be interesting to the check the viability of Natural Inflation in the framework of EGB. The potential for Natural Inflation can be written as, 
 \begin{equation}
 V = V_0 \bigg [1+ \cos \bigg(\frac{\phi}{f}   \bigg) \bigg].
 \end{equation}
 Using the definition of slow roll parameters (\ref{EGBsr1}),(\ref{EGBsr2}), (\ref{EGBsr3}) and (\ref{slrVeffd}) we write

\begin{equation}
\epsilon_1=~- \frac{\xi_1 \xi_2 \sech^2(\xi_2 \phi) \sin \left( \frac{\phi}{f} \right) }{3 f U}+ \frac{U \tan^2 \left( \frac{\phi}{2 f} \right)}{f^2}
\label{natepsiolon1}
\end{equation}
\begin{footnotesize}
\begin{align}
\epsilon_2 =& \frac{1}{3 f^2 U \left( 1+ \cos \left(\frac{\phi}{f} \right) \right)^2} \Bigg\{ 12 U^2  \left( 1+ \cos \left(\frac{\phi}{f} \right) \right)+ 8 f \xi_1 \xi_2 \cos^4\left( \frac{\phi}{2 f} \right) \cot\left( \frac{\phi}{2 f} \right) \sech^2 \left(\xi_2 \phi \right) \nonumber\\&\left[- \cos\left(\frac{\phi}{f} \right)+2 f \xi_2 \sin \left(\frac{\phi}{f} \right)  \tanh \left( \xi_2 \phi \right)\right] \Bigg\}
\end{align}
\begin{align}
\delta_1 = -\frac{1}{9 f U^3} \Bigg\{ 2 \xi_1 \xi_2 \sech^2 \left( \xi_2 \phi \right) \left[ 4 f \xi_1 \xi_2 \cos^4 \left( \frac{\phi}{2f} \right) \sech^2 \left( \xi_2 \phi \right) - 3 U^2 \sin \left( \frac{\phi}{f} \right) \right]   \Bigg\}
\end{align}
\begin{align}
\delta_2 = &\frac{1}{3 f^2 U \left(1+ \cos \left( \frac{\phi}{f}  \right) \right)} \Bigg\{ 2 f \xi_1 \xi_2 \sech^2 \left( \xi_2 \phi \right) \left[ 2\sin \left( \frac{\phi}{f} \right) +\sin \left( \frac{2 \phi}{f} \right) + 16 f \xi_2 \cos^4 \left(\frac{\phi}{2 f} \right) \tanh \left( \xi_2 \phi\right)\right] \nonumber\\& + 6 U^2 \left[ \cos \left( \frac{\phi}{f} \right)- 2 f \xi_2 \sin \left( \frac{\phi}{f} \right)\tanh \left( \xi_2 \phi\right)   \right]  \Bigg\}
\label{natdelta2}
\end{align}
\end{footnotesize}
Using Eq. (\ref{N1}) we can write the expression of number of e-fold for the Natural inflation as:
\begin{align}
N =  \int\limits^{\phi}_{\phi_{end}} \frac{3 fU \left(1+\cos \left(\frac{\phi}{f} \right) \right)}{8 f \xi_1 \xi_2 \cos^4 \left( \frac{\phi}{2 f} \right) \sech^2 \left( \xi_2 \phi  \right) - 6 U^2 \sin \left( \frac{\phi}{f} \right)}
\label{natefold}
\end{align} 
 Again it is difficult to solve the Eq. (\ref{natefold}) analytically, so we follow the same numerical approach as mentioned before. From Eq. (\ref{natepsiolon1}-\ref{natdelta2}) we calculate $r$ and $n_s$ as mentioned in the table \ref{tab3}
 \begin{center}
\begin{table*}[!htb]
\begin{center}
\begin{tabular}{|c|c|c|c|c|c|c|}
\hline
   & \multicolumn{2}{c|}{$f=0.5$} & \multicolumn{2}{c|}{$f=0.7$} & \multicolumn{2}{c|}{$f=0.9$} \\ \hline
$N$  & $n_s$             & $r$            & $n_s$             & $r$           & $n_s$             & $r$               \\ \hline
55 & 0.9502         & 0.03898          &0.9601       & 0.07752      & 0.9625       & 0.10040         \\ \hline
65 & 0.9537        & 0.02505       & 0.9650       & 0.05803      & 0.9678       & 0.07929        \\ \hline
75 & 0.9558        & 0.01642       &0.9684      & 0.04450     &0.9716      & 0.06402         \\ \hline
\end{tabular}
\end{center}
\caption{Values of the inflationary observables $r$ and $n_{s}$ for different values of $f$ and number of e-folds $N$. The observables are in good agreement with the $Planck'18$ \cite{Planck2018} }.
\label{tab3}
\end{table*}
\end{center}

 \begin{center}
\begin{figure}[htb!]
\centering
\includegraphics[scale=0.44]{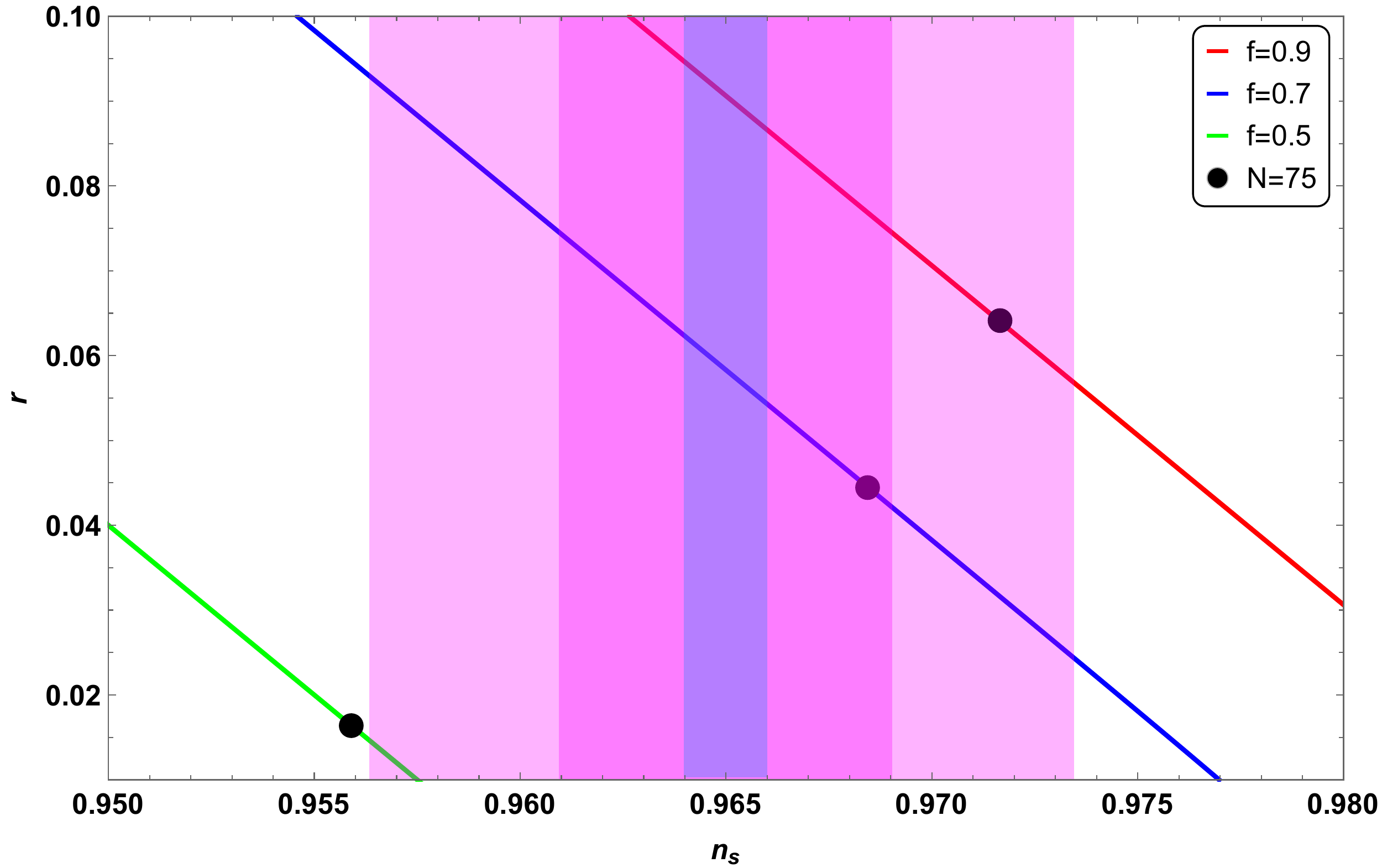}
\caption{Plots of $r$ and $n_s$, for $U=0.005$, for Natural inflation. The light pink shaded region corresponds to $2-\sigma$ and dark pink shaded region corresponds to $1-\sigma$ bounds on $n_s$ from {\it Planck'18}. The deep blue shaded region corresponds to the $1-\sigma$ bounds of future CMB observations  \cite{Euclid,PRISM} keeping the same central value.   }
\label{rnsnatU005}
\end{figure}
\end{center}
\begin{figure}[t]
\centering
\includegraphics[scale=0.22]{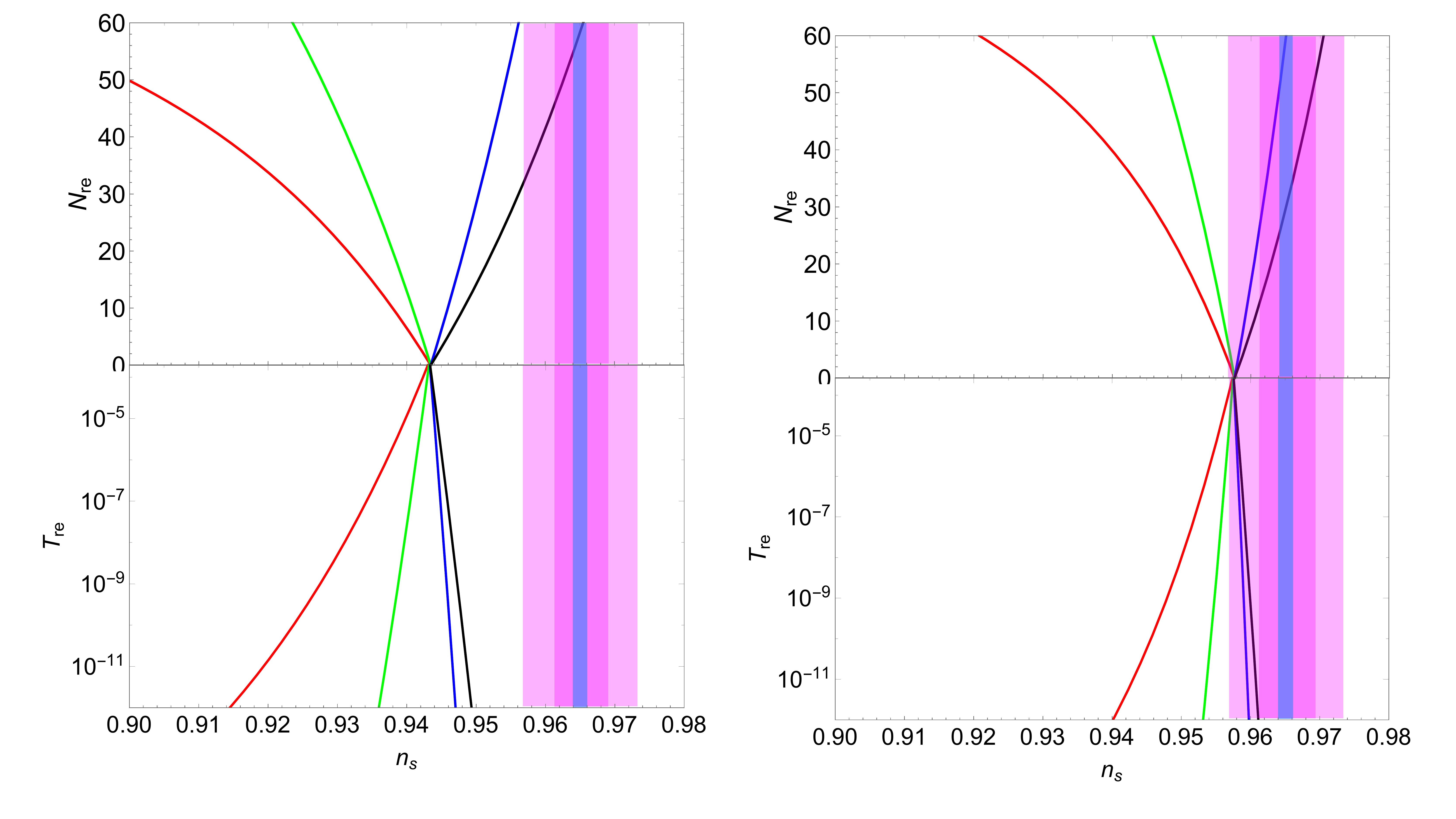}
\caption{ {\small   { \bf Left panel: }Plots for Tre and Nre for Goldstone inflation as a function of  $n_s$ for fixed value of $U=0.5$ and $f=0.7$, four different color corresponds to different values of $\omega$. Red color stands for $\omega= -\frac{1}{3}$, green for $\omega= 0$, blue for $\omega= \frac{2}{3}$ and black for $\omega= 1$. The light pink shaded region corresponds to $2-\sigma$ and dark pink shaded region corresponds to $1-\sigma$ bounds on $n_s$ from {\it Planck'18}. The deep blue shaded region corresponds to the $1-\sigma$ bounds of future CMB observations.  } {\bf Right Panel: }  Plots for Tre and Nre for Goldstone inflation as a function of  $n_s$ for fixed value of $U=0.005$ and $f=0.7$, four different color corresponds to different values of $\omega$. Red color stands for $\omega= -\frac{1}{3}$, green for $\omega= 0$, blue for $\omega= \frac{2}{3}$ and black for $\omega= 1$. The light pink shaded region corresponds to $2-\sigma$ and dark pink shaded region corresponds to $1-\sigma$ bounds on $n_s$ from {\it Planck'18}. The deep blue shaded region corresponds to the $1-\sigma$ bounds of future CMB observations  \cite{Euclid,PRISM} keeping the same central value.}
\label{EGBTreNre}
\end{figure}
\section{Reheating }
\label{reheatingegb}
At the end of inflation, owing to the fact that the universe expands exponentially, it ends up in a supercooled state. So for the universe to reheat itself, to enter the radiation dominated era, and to start the Big Bang Nucleosynthesis(BBN), there is a need for a mechanism through which the universe can come out of this supercooled state. \cite{33, 34, 35, 36, 37, 38, 39}. For other realisations of inflationary dynamics in non standard scenarios e.g. Warm inflation, readers are suggested to go through Ref. \cite{40, 41, 42, 43,44,44a}, where the reheating phase is not required, and after the end of inflation, we go directly into the radiation dominated phase. This transition of the universe, from supercooled state to a hot, thermal and radiation dominated state can be realised either through perturbative process, known as perturbative reheating, or the process of parametric resonance, better known as (p)reheating (For detailed discussion reader is suggested to follow \cite{45}). The epoch of reheating can be parametrized by $N_{re}$ (number of e-folds during the reheating phase), $T_{re}$ (thermalization temperature), and the equation of states during reheating ($w_{re}$) \cite{46, 47}. The analysis here is independent of the exact dynamical process of reheating, and we can still explore the parameter space. \cite{48,Adhikari:2019uaw}. 
\begin{figure}[t]
\centering
\includegraphics[scale=0.25]{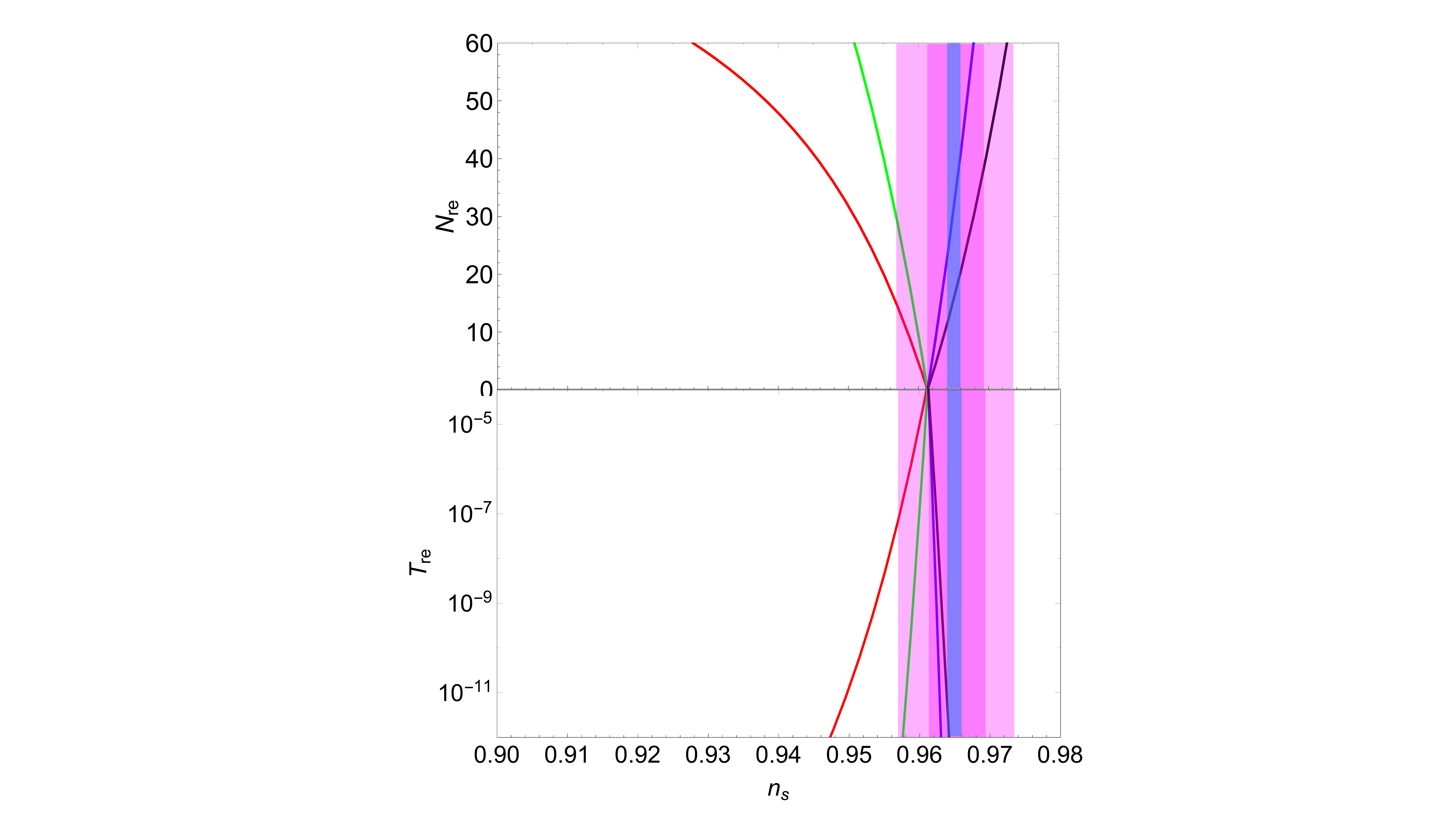}
\caption{ Plots for Tre and Nre for Natural inflation as a function of  $n_s$ for fixed value of $U=0.005$ and $f=0.7$, four different color corresponds to different values of $\omega$. Red color stands for $\omega= -\frac{1}{3}$, green for $\omega= 0$, blue for $\omega= \frac{2}{3}$ and black for $\omega= 1$. The light pink shaded region corresponds to $2-\sigma$ and dark pink shaded region corresponds to $1-\sigma$ bounds on $n_s$ from {\it Planck'18}. The deep blue shaded region corresponds to the $1-\sigma$ bounds of future CMB observations  \cite{Euclid,PRISM} keeping the same central value.  } 
\label{EGBTreNreNat}
\end{figure}

\begin{equation}
N_{re}= \frac{4}{ (1-3w_{re} )}   \left[61.488  - \ln \left(\frac{ V_{end}^{\frac{1}{4}}}{ H_{k} } \right)  - N_{k}   \right]
\label{re7}
\end{equation}

\begin{equation}
T_{re}= \left[ \left(\frac{43}{11 g_{re}} \right)^{\frac{1}{3}}    \frac{a_0 T_0}{k_{}} H_{k} e^{- N_{k}} \left[\frac{3^2 \cdot 5 V_{end}}{\pi^2 g_{re}} \right]^{- \frac{1}{3(1 + w_{re})}}  \right]^{\frac{3(1+ w_{re})}{3 w_{re} -1}}.
\label{re8}
\end{equation}

Here we have used Planck's pivot  ($k$) of order $0.05 \; \mbox{Mpc}^{-1}$ and $g_{re} \approx 100 $. To evaluate $N_{re}$ and $T_{re}$, one need to calculate the $H_{k}$, $N_{k}$ and $V_{end}$ for the given potential. Using the Eq. (\ref{As}) we can establish the relation between $H_k$ and $n_s$, similarly $N_k$ can be written in terms of spectral index ($n_s$) . From $\epsilon=1$ condition at the end of inflation, it is straightforward to calculate $V_{end}$.  Equipped with all the preliminaries and using Eq. (\ref{re7}) and (\ref{re8}) we compute the reheating temperature and number of e-folds during reheating, for the different equation of state ($\omega_{re}$) as shown in the fig. (\ref{EGBTreNre}) and (\ref{EGBTreNreNat}). 
Even though the reheating analysis can be done for any of the values of $f$ that we have done our inflationary analysis on, we have done our calculations for $f=0.7$. From  fig. (\ref{EGBTreNre}) it is evident that if one wants to bound the reheating temperature from the recent CMB observations, lower value of $U$ is preferable while keeping the other parameters $\xi_1$ and $\xi_2$ fixed. 

\section{Conclusion}
\label{egbconclusion}
\noindent
In this paper, we have analysed and studied the validity of inflationary models in the framework of Einstein-Gauss-Bonnet gravity. Due to the presence of the extra Gauss-Bonnet coupling in the action, the inflationary dynamics deviate from the standard case. We tested two inflationary models in EGB: Goldstone Inflation and Natural Inflation. In the case of Goldstone inflation, we calculated the inflationary parameters $r$
and $n_{s}$, for two different values of $U$, and corresponding three different sub-Planckian values of $f$. The results of which are given in Table \ref{tab1} and Table \ref{tab2}. For the case of $U=0.005$, we are able to better constrain the inflationary observables, $r$ and $n_{s}$, using the {\it Planck'18} \cite{Planck2018}. For $U=0.005$ and $f=0.7$, we also studied the phase of reheating, where we use an indirect approach to constrain the reheating parameters, namely the the reheating temperature ($T_{re}$) and the number of e-folds ($N_{re}$). This approach is independent of the exact dynamics of the reheating phase. For $U=0.5$, the reheating parameters can not be constrained through {\it Planck'18} \cite{Planck2018}. We repeated the aforementioned calculations for Natural Inflation. In this case, if we take $U=0.5$, it is difficult to achieve the adequate number of e-folds required for a successful phase of inflation. For $U=0.005$, we were able to perform the calculations and we found that the inflationary observations are in agreement with {\it Planck'18} \cite{Planck2018}. Results have been laid out in Table \ref{tab3}. We were also be to constrain the reheating parameters in this case. While we only picked one particular values of $f (=0.7)$ to study the phase of reheating,it is rather straight forward to check reheating for other values of $f$. 

We found that for both the models of inflation, the inflationary and reheating parameters are in good agreement with {\it Planck'18} \cite{Planck2018}, in the sub-Planckian regime.

Another interesting aspect of the early universe that can be studied in the domain of EGB gravity is the production of Primordial Blackholes \cite{Kawai:2021edk}. The seeds for PBH formation, during inflation, can be provided by an appropriate choice of the form of EGB coupling. It is not possible to probe the scales associated with the seeds of PBH production, so to study the production mechanism of PBHs will have interesting phenomenological implications. We will come back to this in our future studies. 

\section{Acknowledgement}
\noindent
Authors would like to thank Mayukh Raj Gangopadhyay for the initial motivation, and fundamental insights at every stage of this work. Yogesh would like to thank Imtiyaz Ahmad Bhat for the useful discussions. 
 H.A.K. would like to thank Tabish Qureshi and Centre For Theoretical Physics for providing the facilities during the course of this work.


\begin{thebibliography}{99}


\bibitem{croon} D.~Croon, V.~Sanz, J.~Setford, JHEP \textbf{10}, 020 (2015), \href{https://arxiv.org/abs/1503.08097}{[arXiv:1503.08097]}.

\bibitem{Planck2018}
    {Y.~Akrami {\it et al.}} [Planck Collaboration],
   {Planck 2018 results. X. Constraints on inflation},
  \href{https://arxiv.org/abs/1807.06211}{[arXiv:1807.06211].}
  
\bibitem{Bhattacharya:2018xlw}
S.~Bhattacharya and M.~R.~Gangopadhyay,
Phys. Rev. D \textbf{101}, no.2, 023509 (2020)
doi:10.1103/PhysRevD.101.023509
\href{https://arxiv.org/abs/1812.08141}{[arXiv:1812.08141].}

\bibitem{guth} 
  A.~H.~Guth,
  \href{https://journals.aps.org/prd/abstract/10.1103/PhysRevD.23.347}{Phys.\ Rev.\ D {\bf 23}, 347 (1981)}

\bibitem{Liddle}
A. R. Liddle and D. H. Lyth, {\it Cosmological Inflation and Large 
Scale Structure}, (Cambridge University Press: Cambridge, UK), (1998).




\bibitem{cmbinflate}
E. W. Kolb and M. S. Turner, 
{\it The Early Universe}, (Addison-Wesley, Menlo Park, Ca., 1990).
%

\bibitem{Linde:2005ht}A.~D.~Linde,
  Contemp.\ Concepts Phys.\  {\bf 5}, 1 (1990),





\bibitem{Linde:1983gd}
A.D.~Linde,
 {Chaotic Inflation},
Phys. Lett. B \textbf{129}, 177 (1983)


\bibitem{mukha81} V.~F.~Mukhanov, G.~V.~Chibisov, JETP \ Lett. {\bf 33}, 532 (1981).


\bibitem{WMAP9} G. Hinshaw,  et al. ({\it WMAP Collaboration}) ,  Astrophys. J. Suppl. Ser., {\bf 208}, 19 (2013), \href{https://arxiv.org/abs/1212.5226}{[arXiv:1212.5226]}.

\bibitem{PlanckXX}  {\it PlanckXX} Collaboration, Astron. \& Astrophys, {\bf 594}(2016) A20, \href{https://arxiv.org/abs/1502.02114}{[arXiv1502.02114]}.



\bibitem{csaki} C.~Csaki, N.~Kaloper, J.~Serra, J.~Terning, \ Phys. \ Rev. \ Lett.{\bf 113} (2014) 161302, \href{https://arxiv.org/abs/1406.5192}{[arXiv:1406.5192]}.
\bibitem{contino} R.~Contino, Y.~Nomura, A.~Pomarol, \ Nucl. \ Phys. \ B {\bf 671} (2003) 148, \href{https://arxiv.org/abs/hep-ph/0306259}{[arXiv:hep-ph/0306259]}.

\bibitem{freese} K.~Freese, J.~A.~Frieman, A.~V.~Olinto, \ Phys. \ Rev. \ Lett.{\bf 65}, 3233 (1990).  


\bibitem{Adams:1992bn}
F.~C.~Adams, J.~R.~Bond, K.~Freese, J.~A.~Frieman and A.~V.~Olinto,
Phys. Rev. D \textbf{47} (1993), 426-455
\href{https://arxiv.org/abs/hep-ph/9207245}{[arXiv:hep-ph/9207245 [hep-ph]]}.

\bibitem{Kim:2004rp}
J.~E.~Kim, H.~P.~Nilles and M.~Peloso,
JCAP \textbf{01} (2005), 005
doi:10.1088/1475-7516/2005/01/005
\href{https://arxiv.org/abs/hep-ph/0409138}{[arXiv:hep-ph/0409138 [hep-ph]]}.

\bibitem{delaFuente:2014aca}
A.~de la Fuente, P.~Saraswat and R.~Sundrum,
Phys. Rev. Lett. \textbf{114} (2015) no.15, 151303
doi:10.1103/PhysRevLett.114.151303
\href{https://arxiv.org/abs/1412.3457}{[arXiv:1412.3457 [hep-th]]}.


\bibitem{Jensen:1986nf}
L.~G.~Jensen and J.~A.~Stein-Schabes,
Phys. Rev. D \textbf{35} (1987), 1146
doi:10.1103/PhysRevD.35.1146


\bibitem{Freese:2014nla}
K.~Freese and W.~H.~Kinney,
JCAP \textbf{03} (2015), 044
doi:10.1088/1475-7516/2015/03/044
\href{https://arxiv.org/abs/1403.5277}{[arXiv:1403.5277 [astro-ph.CO]]}.



\bibitem{nima} N.~Arkani-Hamed, H.~C.~Cheng, P.~Creminelli, L.~Randall, \ Phys. \ Rev. \ Lett.{\bf90}, 221302 (2003), \href{https://arxiv.org/abs/hep-th/0301218}{[arXiv:hep-th/0301218]}.
\bibitem{lindehyb} A.~D.~Linde, \ Phys. \ Rev. {\bf D49}, 748, (1994),\href{https://arxiv.org/abs/astro-ph/9307002}{ [arXiv:astro-ph/9307002]}.
\bibitem{kim} J.~E.~Kim, H.~P.~Nilles, M.~Peloso, JCAP {\bf 0501}, (2005) 005, \href{https://arxiv.org/abs/hep-ph/0409138}{[arXiv:hep-ph/0409138]}.
\bibitem{nflation1} S.~Dimopoulos, S.~Kachru, J.~McGreevy, J.~G.~Wacker, JCAP {\bf 0808} (2008) 003, \href{https://arxiv.org/abs/hep-th/0507205}{[arXiv:hep-th/0507205]}.
\bibitem{nflation2} A.~R.~Liddle, A.~Mazumdar, F.~E.~Schunck, \ Phys. \ Rev. {\bf D58}, 061301 (1998),\href{https://arxiv.org/abs/astro-ph/9804177}{[arXiv:astro-ph/9804177]}.
\bibitem{nflation3} E.~J.~Copeland, A.~Mazumdar, N.~J.~Nunes, \ Phys. \ Rev. {\bf D60}, 083506 (1999),\href{https://arxiv.org/abs/astro-ph/9904309}{[arXiv:astro-ph/9904309]}.
\bibitem{axionmono} E.~Silverstein, A.~Westphal, \ Phys. \ Rev. {\bf D78}, 106003 (2008),\href{https://arxiv.org/abs/0803.3085}{[arXiv:0803.3085]}.
\bibitem{pngb} N.~Arkani-Hamed, H.~C.~Cheng, P.~Creminelli, L.~Randall, JCAP {\bf 0307}, 003 (2003), \href{https://arxiv.org/abs/hep-th/0302034}{[arXiv:hep-th/0302034]}.

 \bibitem{Kallosh:2013pby}
R.~Kallosh and A.~Linde,
 {Superconformal generalization of the chaotic inflation model $\frac{\lambda}{4} \phi^{4} - \frac{\xi}{2} \phi^{2}R$},
J. Cosmol. Astropart. Phys. \textbf{06}, 027 (2013)
\href{https://arxiv.org/abs/1306.3211}{[arXiv:1306.3211]}. 
  
\bibitem{predictions}
  	A.A. Starobinsky,  {Dynamics of phase transition in the new inflationary universe scenario and generation of perturbations},
		Phys.\ Lett.\ B {\bf 117},  175 (1982).

  A.~Starobinsky,  {The Perturbation Spectrum Evolving from a Nonsingular
Initially de Sitter Cosmology and the Microwave Background Anisotropy}, Sov.\ Astron.\ Lett.\ {\bf{9}}, 302
(1983).

\bibitem{Barvinsky:1994hx}
 		 A.O.~Barvinsky and A.Yu.~Kamenshchik,
  		 {Quantum scale of inflation and particle physics of the early universe,}
 		 {Phys.\ Lett.\ B} {\bf 332}, 270 (1994)
  		\href{https://arxiv.org/abs/gr-qc/9404062}{[arXiv:gr-qc/9404062]}.


\bibitem{Cervantes-Cota1995}
J.L.~Cervantes-Cota and H.~Dehnen,
{Induced gravity inflation in the standard model of particle physics,}
Nucl.\ Phys. B {\bf 442}, 391 (1995) 
\href{https://arxiv.org/abs/astro-ph/9505069}{[arXiv:astro-ph/9505069].}

\bibitem{BezrukovShaposhnikov}
   F.L.~Bezrukov and M.~Shaposhnikov, The Standard
 {Model Higgs boson as the inflaton},
Phys.\ Lett.\ B {\bf 659}, 703 (2008)
 \href{https://arxiv.org/abs/0710.3755}{[arXiv:0710.3755]};\\
 A.O. Barvinsky, A.Y. Kamenshchik and A.A. Starobinsky,
		 {Inflation scenario via the Standard Model Higgs boson and LHC},		
		J. Cosmol. Astropart. Phys. {\bf 0811} 021 (2008)
\href{https://arxiv.org/abs/0809.2104}{[arXiv:0809.2104]};\\
A.~De Simone, M.P.~Hertzberg and F.~Wilczek,
  		 {Running Inflation in the Standard Model,}
 		 {Phys.\ Lett. B}~{\bf 678}, 1 (2009)
 		\href{https://arxiv.org/abs/0812.4946}{[arXiv:0812.4946]};\\	
		F.L.~Bezrukov, A.~Magnin and M.~Shaposhnikov,
		 {Standard Model Higgs boson mass from inflation,}
		 {Phys. Lett. B} {\bf 675}, 88 (2009)
	\href{https://arxiv.org/abs/0812.4950}{[arXiv:0812.4950]};\\
		A.O. Barvinsky, A.Yu.~Kamenshchik, C. Kiefer, A.A. Starobinsky and C.F.~Steinwachs,
		 {Higgs boson, renormalization group, and cosmology,}
		 {Eur. Phys. J.} C \textbf{72}, 2219 (2012)
		\href{https://arxiv.org/abs/0910.1041}{[arXiv:0910.1041]};\\
F.L.~Bezrukov, A.~Magnin, M.~Shaposhnikov and S.~Sibiryakov,
 {Higgs inflation: consistency and generalisations},
JHEP {\bf 1101} 016 (2011)
\href{https://arxiv.org/abs/1008.5157}{[arXiv:1008.5157]};\\	
F.L.~Bezrukov,
		 {The Higgs field as an inflaton,}
		 {Class. Quant. Grav.}~{\bf 30}, 214001 (2013)
		\href{https://arxiv.org/abs/1307.0708}{[arXiv:1307.0708]};\\
J.~Rubio, {Higgs inflation},
Front. Astron. Space Sci. \textbf{5}, 50 (2019)
\href{https://arxiv.org/abs/1807.02376}{[arXiv:1807.02376]}.

		
\bibitem{EOPV2014}
E.~Elizalde, S.D.~Odintsov, E.O.~Pozdeeva, and S.Yu.~Vernov,
		 {Renormalization-group inflationary scalar electrodynamics and $SU(5)$ scenarios confronted with Planck2013 and BICEP2 results,}
		Phys. Rev. D \textbf{90},  084001 (2014)
		\href{http://de.arxiv.org/abs/1408.1285}{[arXiv:1408.1285]}.

\bibitem{Koshelev:2020xby}
A.S.~Koshelev, K.S.~Kumar and A.A.~Starobinsky,
 {Analytic infinite derivative gravity, $R^2$-like inflation, quantum gravity and CMB},
\href{https://arxiv.org/abs/2005.09550}{[arXiv:2005.09550].}

  \bibitem{vandeBruck:2015gjd}
  C.~van de Bruck and C.~Longden,
   {Higgs Inflation with a Gauss-Bonnet term in the Jordan Frame},
  Phys.\ Rev.\ D {\bf 93},  063519 (2016)
 \href{https://arxiv.org/abs/1512.04768}{[arXiv:1512.04768]}.
  
  \bibitem{Guo:2009uk}
  Z.K.~Guo and D.J.~Schwarz,
   {Power spectra from an inflaton coupled to the Gauss-Bonnet term,}
  Phys.\ Rev.\ D {\bf 80}, 063523 (2009)
  \href{https://arxiv.org/abs/0907.0427}{[arXiv:0907.0427].}

\bibitem{Guo}
  Z.K.~Guo and D.J.~Schwarz,
  {Slow-roll inflation with a Gauss-Bonnet correction},
  Phys.\ Rev.\ D {\bf 81}, 123520 (2010)
 \href{https://arxiv.org/abs/1001.1897}{[arXiv:1001.1897].}
  

  
\bibitem{Chakraborty:2018scm}
S.~Chakraborty, T.~Paul and S.~SenGupta,
 {Inflation driven by Einstein-Gauss-Bonnet gravity},
Phys. Rev. D \textbf{98}, 083539 (2018)
\href{https://arxiv.org/abs/1804.03004}{[arXiv:1804.03004].  }
  
  \bibitem{Koh:2016abf}
S.~Koh, B.H.~Lee and G.~Tumurtushaa,
Reconstruction of the Scalar Field Potential in Inflationary Models with a Gauss-Bonnet term,
Phys. Rev. D \textbf{95},  123509 (2017)
\href{https://arxiv.org/abs/1610.04360}{[arXiv:1610.04360].}
  
 \bibitem{Pozdeeva:2020shl}
E.~Pozdeeva,
 Generalization of cosmological attractor approach to Einstein-Gauss-Bonnet gravity,
Eur. Phys. J. C \textbf{80},  612 (2020)
\href{https://arxiv.org/abs/2005.10133}{[arXiv:2005.10133].} 
  
  
  
  
  
  \bibitem{Soda2008}
M.~Satoh and J.~Soda,
 {Higher Curvature Corrections to Primordial Fluctuations in Slow-roll Inflation},
J. Cosmol. Astropart. Phys. \textbf{09}, 019 (2008)
\href{https://arxiv.org/abs/0806.4594}{[arXiv:0806.4594]}.





\bibitem{Jiang:2013gza}
  P.X.~Jiang, J.W.~Hu and Z.K.~Guo,
   {Inflation coupled to a Gauss-Bonnet term},
  Phys.\ Rev.\ D {\bf 88}, 123508 (2013)
 \href{https://arxiv.org/abs/1310.5579}{ [arXiv:1310.5579].}

\bibitem{Koh}
S.~Koh, B.H.~Lee, W.~Lee, and G.~Tumurtushaa,
{Observational constraints on slow-roll inflation coupled to a Gauss-Bonnet term}, Phys. Rev. D \textbf{90},  063527 (2014)
\href{https://arxiv.org/abs/1404.6096}{[arXiv:1404.6096]};\\
S.~Koh, B.H.~Lee and G.~Tumurtushaa,
Constraints on the reheating parameters after Gauss-Bonnet inflation from primordial gravitational waves,
Phys. Rev. D \textbf{98}, 103511 (2018)
\href{https://arxiv.org/abs/1807.04424}{[arXiv:1807.04424].}




\bibitem{JoseMathew}
J.~Mathew, S.~Shankaranarayanan,
Low scale Higgs inflation with Gauss-Bonnet coupling,
Astroparticle Physics \textbf{84}, 1 (2016)
\href{https://arxiv.org/abs/1602.00411}{[arXiv:1602.00411].}

\bibitem{vandeBruck:2016xvt}
  C.~van de Bruck, C.~Longden, and  K.~Dimopoulos
 {Reheating in Gauss-Bonnet-coupled inflation,}
  Phys.\ Rev.\ D {\bf 94}, 023506 (2016)
 \href{https://arxiv.org/abs/1605.06350}{ [arXiv:1605.06350].}


\bibitem{Pozdeeva:2020apf}
E.~O.~Pozdeeva, M.~R.~Gangopadhyay, M.~Sami, A.~V.~Toporensky and S.~Y.~Vernov,
Phys. Rev. D \textbf{102} (2020) no.4, 043525
doi:10.1103/PhysRevD.102.043525
\href{https://arxiv.org/abs/2006.08027}{[arXiv:2006.08027].}


\bibitem{Pozdeeva:2016cja}
  E.O.~Pozdeeva, M.A.~Skugoreva, A.V.~Toporensky, and S.Yu.~Vernov,
   {Possible evolution of a bouncing universe in cosmological models with nonminimally coupled scalar fields},
  J. Cosmol. Astropart. Phys. {\bf 1612}, no. 12, 006 (2016)
\href{https://arxiv.org/abs/1608.08214}{[arXiv:1608.08214].}

\bibitem{Nozari:2017rta}
  K.~Nozari and N.~Rashidi,
   {Perturbation, nonGaussianity, and reheating in a Gauss-Bonnet $\alpha$-attractor model},
  Phys.\ Rev.\ D {\bf 95}, 123518 (2017)
  \href{https://arxiv.org/abs/1705.02617}{[arXiv:1705.02617].}

\bibitem{Armaleo:2017lgr}
J.M.~Armaleo, J.~Osorio Morales and O.~Santillan,
Gauss-Bonnet models with cosmological constant and non zero spatial curvature in $D=4$,
Eur. Phys. J. C \textbf{78}, 85 (2018)
\href{https://arxiv.org/abs/1711.09484}{[arXiv:1711.09484].}




\bibitem{Yi:2018gse}
Z.~Yi, Y.~Gong, and M.~Sabir,
Inflation with Gauss-Bonnet coupling,
Phys. Rev. D \textbf{98},  083521 (2018)
\href{https://arxiv.org/abs/1804.09116}{[arXiv:1804.09116]};\\
Z.~Yi and Y.~Gong,
 {Gauss-Bonnet Inflation and the String Swampland},
Universe \textbf{5}, no.9, 200 (2019)
\href{https://arxiv.org/abs/1811.01625}{[arXiv:1811.01625].}



\bibitem{Odintsov:2018}
   S.D.~Odintsov and V.K.~Oikonomou,
   {Viable Inflation in Scalar-Gauss-Bonnet Gravity and Reconstruction from Observational Indices},
  Phys.\ Rev.\ D {\bf 98}, 044039 (2018)
  \href{https://arxiv.org/abs/1808.05045}{[arXiv:1808.05045]};\\
S.~Nojiri, S.~Odintsov, V.~Oikonomou, N.~Chatzarakis and T.~Paul,
 {Viable inflationary models in a ghost-free Gauss-Bonnet theory of gravity},
Eur. Phys. J. C \textbf{79}, no.7, 565 (2019)
\href{https://arxiv.org/abs/1907.00403}{[arXiv:1907.00403].}

\bibitem{Fomin:2019yls}
  I.V.~Fomin and S.V.~Chervon,
   {Reconstruction of GR cosmological solutions in modified gravity theories},
  Phys. Rev. D \textbf{100}, 023511 (2019) [arXiv:1903.03974];\\
I.~Fomin,
 {Gauss-Bonnet term corrections in scalar field cosmology},
\href{https://arxiv.org/abs/2004.08065}{[arXiv:2004.08065].}


\bibitem{Kleidis:2019ywv}
K.~Kleidis and V.~Oikonomou,
 {A Study of an Einstein Gauss-Bonnet Quintessential Inflationary Model},
Nucl. Phys. B \textbf{948}, 114765 (2019)
\href{https://arxiv.org/abs/1909.05318}{[arXiv:1909.05318].}

\bibitem{Rashidi:2020wwg}
N.~Rashidi and K.~Nozari,
 {Gauss-Bonnet Inflation after Planck2018},
Astrophys. J. \textbf{890}, 58 (2020)
\href{https://arxiv.org/abs/2001.07012}{[arXiv:2001.07012].}

\bibitem{Odintsov:2020sqy}
S.~Odintsov, V.~Oikonomou and F.~Fronimos,
 {Rectifying Einstein-Gauss-Bonnet Inflation in View of GW170817},
\href{https://arxiv.org/abs/2003.13724}{[arXiv:2003.13724]};\\
S.~Odintsov and V.~Oikonomou,
 {Swampland Implications of GW170817-compatible Einstein-Gauss-Bonnet Gravity},
Phys. Lett. B \textbf{805}, 135437 (2020)
\href{https://arxiv.org/abs/2004.00479}{[arXiv:2004.00479].}
  
\bibitem{Kawai:2021bye}
        S.~Kawai and J.~Kim,
        Phys. Rev. D \textbf{104}, no.4, 043525 (2021)
        \href{https://arxiv.org/abs/2105.04386}{[arXiv:2105.04386].}
        
        
 \bibitem{Kawai:2017kqt}
        S.~Kawai and J.~Kim,
        Phys. Lett. B \textbf{789}, 145-149 (2019)
        \href{https://arxiv.org/abs/1702.07689}{[arXiv:1702.07689].}  
  
  
  
  
  
\bibitem{Pozdeeva:2019agu}
  E.O.~Pozdeeva, M.~Sami, A.V.~Toporensky and S.Yu.~Vernov,
   {Stability analysis of de Sitter solutions in models with the Gauss-Bonnet term},
  Phys.\ Rev.\ D {\bf 100},  083527 (2019)
  \href{https://arxiv.org/abs/1905.05085}{[arXiv:1905.05085].}
  
\bibitem{Kawai:2021edk}
        S.~Kawai and J.~Kim,
        Phys. Rev. D \textbf{104}, no.8, 083545 (2021)
       \href{https://arxiv.org/abs/2108.01340}{ [arXiv:2108.01340].}
  
\bibitem{Hikmawan:2015rze}
G.~Hikmawan, J.~Soda, A.~Suroso, and F.P.~Zen,
Comment on ''Gauss-Bonnet inflation'',
Phys. Rev. D \textbf{93}, 068301 (2016)
\href{https://arxiv.org/abs/1512.00222}{[arXiv:1512.00222]. }
  

\bibitem{Mukhanov:2013tua}
V.~Mukhanov,
 {Quantum Cosmological Perturbations: Predictions and Observations},
Eur. Phys. J. C \textbf{73}, 2486 (2013)
\href{https://arxiv.org/abs/1303.3925}{[arXiv:1303.3925].}


\bibitem{PRISM}
P.~Andre \textit{et al.} [PRISM],
\href{https://arxiv.org/abs/1306.2259}{[arXiv:1306.2259].}  
  
  
 \bibitem{Euclid}
L.~Amendola \textit{et al.} [Euclid Theory Working Group],
Living Rev. Rel. \textbf{16} (2013), 6
doi:10.12942/lrr-2013-6
\href{https://arxiv.org/abs/1206.1225}{[arXiv:1206.1225]. }
  

 \bibitem{33} 
  V.~F.~Mukhanov, H.~A.~Feldman and R.~H.~Brandenberger, Phys.\ Rept.\  {\bf 215}, 203 (1992).
 
 \bibitem{34}   
 Andreas Albrecht, Paul J. Steinhardt, Michael S. Turner, and Frank Wilczek, Phys. Rev. Lett. {\bf 48}  (1982)
 

 \bibitem{35} 
  L.~Kofman, A.~D.~Linde and A.~A.~Starobinsky, Phys.\ Rev.\ Lett.\  {\bf 73}, 3195 (1994); \href{https://arxiv.org/abs/hep-th/9405187}{[arXiv:9405187].}
  
 
\bibitem{36} 
  Y.~Shtanov, J.~H.~Traschen and R.~H.~Brandenberger, Phys.\ Rev.\ D {\bf 51}, 5438 (1995); \href{https://arxiv.org/abs/hep-ph/9407247}{[arXiv:9407247].}
  
  

\bibitem{37} 
  L.~Kofman, A.~D.~Linde and A.~A.~Starobinsky, Phys.\ Rev.\ D {\bf 56}, 3258 (1997); 
  \href{https://arxiv.org/abs/hep-ph/9704452}{[arXiv:9704452].}
  
  

\bibitem{38} 
  B.~A.~Bassett, S.~Tsujikawa and D.~Wands, Rev.\ Mod.\ Phys.\  {\bf 78}, 537 (2006); 
  \href{https://arxiv.org/abs/astro-ph/0507632}{[arXiv:0507632].}
  
 

\bibitem{39} 
  T.~Rehagen and G.~B.~Gelmini, JCAP {\bf 1506}, no. 06, 039 (2015); 
  \href{https://arxiv.org/abs/1504.03768}{[arXiv:1504.03768].}
  
 

\bibitem{40} 
  A.~Berera, Phys.\ Rev.\ Lett.\ {\bf 75}, 3218 (1995); \href{https://arxiv.org/abs/astro-ph/9509049}{[arXiv:9509049]}.


\bibitem{41} 
  A.~Berera and L.~Z.~Fang, Phys.\ Rev.\ Lett.\  {\bf 74}, 1912 (1995); 
  \href{https://arxiv.org/abs/astro-ph/9501024}{[arXiv:9501024].}
  
 
  

\bibitem{42} 
  M.~Bastero-Gil et. al., Phys.\ Rev.\ Lett.\  {\bf 117}, no. 15, 151301 (2016); 
  \href{https://arxiv.org/abs/1604.08838}{[arXiv:1604.08838].}
 
 
  
 
\bibitem{43} 
  M.~Bastero-Gil et. al., JCAP {\bf 1802}, 054 (2018); 
  \href{https://arxiv.org/abs/1710.10008}{[arXiv:1710.10008].}
  %
  
  
  

\bibitem{44}
M.~R.~Gangopadhyay, $et,al,$ Phys. Rev. D \textbf{103}, no.4, 043505 (2021); 
\href{https://arxiv.org/abs/2011.09155}{[arXiv:2011.09155].}

\bibitem{44a}
S.~Basak, S.~Bhattacharya, M.~R.~Gangopadhyay, N.~Jaman, R.~Rangarajan and M.~Sami,
\href{https://arxiv.org/abs/2110.00607}{[arXiv:2110.00607].}

  \bibitem{45} 
  K.~D.~Lozanov;  \href{https://arxiv.org/abs/1907.04402}{[arXiv:1907.04402].}
  
  
 
\bibitem{46} 
  J.~Martin, C.~Ringeval and V.~Vennin, Phys.\ Rev.\ Lett.\  {\bf 114}, no. 8, 081303 (2015); 
 \href{https://arxiv.org/abs/1410.7958}{[arXiv:1410.7958].}
  
  
 
\bibitem{47} 
  R.~C.~de Freitas and S.~V.~B.~Gonçalves; 
  \href{https://arxiv.org/abs/1509.08500}{[arXiv:1509.08500].}
  
  
 
\bibitem{48} 
  J.~L.~Cook, E.~Dimastrogiovanni, D.~A.~Easson and L.~M.~Krauss, JCAP {\bf 1504}, 047 (2015); \href{https://arxiv.org/abs/1502.04673}{[arXiv:1502.04673].}



\bibitem{Adhikari:2019uaw}
R.~Adhikari, M.~R.~Gangopadhyay and Yogesh,
\href{https://arxiv.org/abs/1909.07217}{[arXiv:1909.07217].}

\end{thebibliography}
\end{document}